# Determination of the spin orientation of helical electrons in monolayer $WTe_2$

Cheng Tan[1,#], Ming-Xun Deng[2,#], Feixiang Xiang[3], Guolin Zheng[1], Sultan Albarakati[1], Meri Algarni[1], James Partridge[1], Alex R. Hamilton[3,*], Rui-Qiang Wang[2,*] and Lan Wang[1,*]

[1]School of Science and ARC Centre of Excellence in Future Low-Energy Electronics Technologies, RMIT Node, RMIT University, Melbourne, VIC 3000, Australia.

[2]Guangdong Provincial Key Laboratory of Quantum Engineering and Quantum Materials, SPTE, South China Normal University, Guangzhou 510006, China.

[3]School of Physics and ARC Centre of Excellence in Future Low-Energy Electronics Technologies, UNSW Node, University of New South Wales, Sydney, NSW 2052, Australia.

[#] These authors equally contribute to the paper.

[*] Corresponding authors. Correspondence and requests for materials should be addressed to A. H. (email: alex.hamilton@unsw.edu.au), R-Q. W. (email: wangruiqiang@m.scnu.edu.cn) and L. W. (email: lan.wang@rmit.edu.au).

## Abstract

**Monolayer $WTe_2$ is predicted to be a quantum spin Hall insulator (QSHI) and electron transport along its edges has been experimentally observed. However, the 'smoking gun' of QSHI, spin-momentum locking of the edge electrons, has not been experimentally demonstrated. We propose a model to establish the relationship between the anisotropic magnetoresistance (AMR) and spin orientation of the helical electrons in $WTe_2$. Based on the predictions of the model, angular dependent magnetoresistance measurements were carried out. The experimental results fully supported the model and the spin orientation of the helical edge electrons was determined. Our results not only demonstrate that $WTe_2$ is indeed a QSHI, but also suggest a convenient method to determine the spin orientation of other QSHIs.**

**Maintext**

The quantum spin hall effect has attracted widespread attention since its theoretical prediction and experimental demonstration (1-4). A quantum spin Hall insulator (QSHI), which can be regarded as a time reversal symmetric topological insulator (TI) in two dimensions (2D), has a one-dimensional (1D) conductive edge mode and insulating bulk states when the Fermi level lies in its bulk band gap. In a QSHI, the electrons are only transported along the edge channels and their spin and momentum are locked, which means the spin orientation of electrons is determined by their moment direction. QSHI was firstly realized on HgTe/CdTe (5-7) and InAs/GaSb (8-10) quantum wells at liquid Helium temperatures. Later, monolayer $WTe_2$ was predicted (11) to be a QSHI with a much larger band gap (0.5 eV) compared with that of HgTe/CdTe and InAs/GaSb, a significant finding since this can increase the quantum spin Hall critical temperature. Soon after the theoretical prediction, the existence of the conductive edge states up to 100 K was confirmed experimentally in monolayer $WTe_2$ (12-14). This has stimulated extensive research on monolayer $WTe_2$. Important physical phenomena have been observed in mono- and few-layer $WTe_2$, such as superconductivity (15, 16), ferroelectricity (17), photo galvanic effect (18), non-linear Hall effect (19, 20), *etc*. However, the basic property of a QSHI, spin-momentum locking, has yet to be experimentally demonstrated in monolayer $WTe_2$. Without the confirmation of spin momentum locking, the question remains as to whether monolayer $WTe_2$ is truly a QSHI. Finding a method to confirm the spin momentum locking and determine the spin orientation is essential for the research of QSHI.

In ferromagnets and materials with strong spin orbit coupling, the probabilities of electron scattering to different directions are related to the spin orientations. Hence, anisotropic magnetoresistance (AMR) can be used as an effective method to probe the spin orientation. For example, the in-plane AMR can be used to determine the sign of the spin polarization in ferromagnetic metals (21, 22). Positive AMR means positive spin polarization (more spin up electrons on the Fermi level) and vice versa. The spin polarized electrons induced by the spin momentum locking effect can also generate the AMR effect, which has been demonstrated both experimentally and theoretically (23-25). In order to use AMR as a tool to confirm the spin momentum locking in monolayer $WTe_2$, we performed both theoretical modelling and experimental measurements. Our model demonstrates that the AMR in $WTe_2$ is determined by the relative direction of the applied magnetic field and the spin orientation of the edge transport electrons. The spins of the edge transport electrons lie in the plane perpendicular to the current direction and sample surface and point to a direction where the lowest AMR is shown. Based on the theory, we performed AMR measurements on h-BN gated monolayer $WTe_2$ devices. The measurement results agree with the predictions of the model, which confirm the spin momentum locking in monolayer $WTe_2$ and its identity as a QSHI.

In-plane and out-of-plane AMR have been widely investigated in the surface states of TIs. Generally, the origin of the in-plane AMR in the TI's surface state is always discussed in relation to the planar Hall effect (PHE). In the report of Taskin *et al.*(23), the PHE from the surface of the $Bi_{2-x}Sb_xTe_3$ film is attributed to electron scattering from magnetic impurities polarized by an in-plane magnetic field. Additionally, Sulaev *et al.* (24) reported the sign change of AMR and attributed it to the in-plane magnetic field induced shift of two coupled surface states. Very recently, based on the previous experimental results, we proposed a microscopic mechanism (25) to explain the AMR and PHE on the surface of a 3D TI. In this theory, the applied magnetic field induces a tilted Dirac cone, which results in anisotropic backscattering and hence generates the AMR and PHE on the surface of 3D TIs. While our previous work focuses on 2D topological surface state in 3D TIs, we can expand this model to 1D edge states in QSHIs, where the Dirac cone turns into 2 crossed-lines, to reveal the relationship between the observed AMR and the spin orientation of electrons in the edge states in monolayer $WTe_2$. To better understand the relationship between the AMR and the spin orientation locked with momentum in monolayer $WTe_2$, we start with a simple effective four-band model for the $WTe_2$ monolayer (26), whose Hamiltonian is $H^{\mathrm{eff}}(\vec{k}) = H_0(\vec{k}) + H_1(\vec{k})$ with

$$H_0(\mathbf{k}) = \varepsilon_{\mathbf{k}} + m_{\mathbf{k}}\tau_z + \hbar(v_x k_x \sigma_z \tau_x - v_y k_y \tau_y) \quad (1)$$

describing the electronic behavior for unperturbed monolayer $WTe_2$ and

$$H_1(\mathbf{k}) = \sigma_z(\lambda k_y - \delta_z \tau_y) + \alpha_x k_x \sigma_y + \alpha_y k_y \sigma_x - \delta_x \sigma_y \tau_x \quad (2)$$

capturing the vertical electric-field-induced spin-orbit coupling allowed by the symmetry of monolayer $WTe_2$, in which $\varepsilon_{\mathbf{k}} = c_0 + c_1 k_x^2 + c_2 k_y^2$, $m_{\mathbf{k}} = m_0 - m_1 k_x^2 - m_2 k_y^2$ and $\sigma_{i=x,y,z}$ ( $\tau_i$ ) stands for Pauli matrix for the spin (orbit) degree of freedom. The spin-orbit coupling induced by the electric field is usually much weaker than the atomic spin-spin coupling, such that we can treat $H_1(\mathbf{k})$ as a small perturbation to the pristine part $H_0(\mathbf{k})$. Provided the model Hamiltonian is defined on the region $-L/2 < y < L/2$ in the x-y plane, we can derive an effective Hamiltonian for the edge states ( see the Supplementary ),

$$H_{edge}^{\beta} = \gamma k_x^2 + \hbar k_x \begin{pmatrix} -v_F & -iv_\alpha \\ iv_\alpha & v_F \end{pmatrix} = \gamma k_x^2 + \hbar k_x \vec{\sigma} \cdot \mathbf{n} \quad (3)$$

with $\vec{\sigma} = (\sigma_x, \sigma_y, \sigma_z)$ and $\mathbf{n} = (0, v_\alpha, -v_F)$, where $\gamma = c_1 - m_1 c_2 / m_2$, $v_F = \beta v_x \sqrt{1 - c_2^2 / m_2^2}$ and $\beta = \pm$ for the top/bottom edge. Then, the spin polarization of the edge electrons can be obtained as $\vec{s} = \langle \vec{\sigma} \rangle = \eta(0, \sin\theta_p, -\cos\theta_p)$, with $\theta_p = \tan^{-1}(v_\alpha / v_F)$ and $\eta = \pm 1$ for the conduction/valence band. As can be seen, the spin of the edge electrons, locked to $k_x$, lies in the y-z plane (perpendicular to the current direction) and is polarized along $\mathbf{n}$, as sketched in Fig.1(a).

In order to determine the spin orientation $\mathbf{n}$, we can calculate the in-plane AMR and out-of-plane AMR by rotating an applied magnetic field $\mathbf{B}$, respectively, as illustrated in upper and down schematics in Fig. 1(a). When $\mathbf{B} = B(\sin\theta\cos\phi, \sin\theta\sin\phi, \cos\theta)$ is added, it interacts with the edge electrons via the Zeeman effect described by $H_Z = -g\mu_B\vec{\sigma}\cdot\mathbf{B}$, where $\mu_B$ is the Bohr magnetron and $g$ represents the g-factor for the edge electrons. As in the 2D topological surface of Ref. (25) the in-plane magnetic field will tilt the Dirac cone of edge states towards x-axis, which lifts the backscattering prohibited by the spin-momentum locking and so enhances the in-plane resistance. With the magnetic field rotating in the sample surface, the dispersion's tilt changes and the backscattering is alternately enhanced and suppressed and then modulates the resistance.

By employing the method of the Boltzmann equation (see the Supplementary ), at low temperatures, we can derive the resistivity for a single edge channel as

$$\rho(\theta,\phi) = \frac{h}{e^2}\frac{L}{v_F\tau_a}\left(\sum_{\chi=\pm}\left|\frac{\hbar v_F k_\chi}{\sqrt{\left(\hbar v_F k_\chi\right)^2 + M_x^2 + M_y^2}} + 2\frac{\gamma}{\hbar v_F}\frac{M_z}{\hbar v_F}\right|\right)^{-1}, \quad (4)$$

with $k_\chi = \left(\chi\sqrt{\frac{1}{(1-\xi^2)^2} - \frac{M_x^2 + M_y^2}{E_F^2(1-\xi^2)}} - \frac{\xi}{1-\xi^2}\right)\frac{E_F}{\hbar v_F}$, where $\xi = 2\gamma M_z/(\hbar v_F)^2$,

$v_F = \sqrt{v_\alpha^2 + v_F^2}$, $\mathbf{M} = g\mu_B\mathbf{B}$, $M_y = M_y\cos\theta_p + M_z\sin\theta_p$, $M_z = M_y\sin\theta_p - M_z\cos\theta_p$, and the Fermi energy $E_F$ is measured from the Dirac point of the edge spectrum.

The numerical calculations for the AMR are carried out by setting $\theta_p = \pi/5$ (the spin orientation angle is $\pi - \theta_p$) in y-z plane, and the results are shown in Figs. 1(b)-(d). The in-plane AMR in Fig. 1(b) as a function of $\phi$ and the out-of-plane AMR in Fig. 1(c) as a function of $\theta$ are plotted for different magnetic field strengths. Obviously, the in-plane AMR is minimum when the magnetic field is collinear with the y-axis and the out-of-plane AMR is minimum for the magnetic field direction parallel to the spin polarization orientation $\theta = \pi - \theta_p$, as expected above. This implies that one can determine the spin orientation of the edge states by first finding $\phi = \phi_p$ at the minimum of the in-plane AMR, e.g., $\phi_p = \pi / 2$ in Fig. 1(b), and then finding $\theta_p$ from the minimum of the out-of-plane AMR at $\phi = \phi_p$, such as in Fig. 1(c). If $\phi$ is not located at the minimum in-plane AMR, i.e., $\phi \neq \phi_p$, the finding $\theta_p$ from the out-of-plane AMR will deviate from the realistic value and the corresponding AMR is not minimum, as shown in Fig. 1(d).

To experimentally validate the proposed model, we fabricated monolayer $WTe_2$ devices and performed extensive angular dependent magnetoresistance measurements on monolayer $WTe_2$. The first important step was to confirm the 1D edge transport in our $WTe_2$ devices. In total, 7 monolayer $WTe_2$ devices (MWT1, MWT2, … MWT7) were fabricated using a 'pick-up' method (27), the details of these devices are shown in Supplementary material C. All the devices contain bottom electrodes consisting of thin Pt (10-15 nm) supported on a h-BN layer (15 ~ 30 nm), a top graphite gate over the top h-BN dielectric (10~20 nm) and a monolayer $WTe_2$ flake encapsulated by two h-BN structures above. An AC current of 10 nA was applied in standard 2 or 4 terminal transport measurements. The optical image of one $WTe_2$ device (MWT1) is shown in Fig 2a. Fig. 2b shows the gating voltage ($V_g$) dependent conductance (G) curves at various temperatures. The conductance shows broad plateaus at lower temperatures in a certain voltage regime around 0 V. Out of this plateau regime, the conductance increases sharply with the increase of magnitude of both positive and negative voltage, which agrees well with the former results (13, 14). Moreover, the gate voltage dependent conductance at 2 K shows that the conductance values reach ~1000 μS at $V_g = 4$ V, which is higher than in the previous reports, indicating high mobility in a clean device. Fig. 2c, 2e and 2f illustrate the magnetoresistance of MWT1 at 2 K at gate voltages of +4 V, -2.6 V and -6 V, respectively. Based on the gate voltage dependent conductance results (Fig. 2b), we know that + 4 V and -6 V correspond to bulk transport, while the device exhibits 1D edge transport at -2.6 V. As shown in Fig 2c, 2e and 2f, the magnetoresistance characteristics of MWT1 differ significantly at these three gate voltages. The magnetoresistances at both $V_g$ = 4 V and -6 V are small (< 10% at 9 T field). In the low field regime, the magnetoresistance only shows weak anti-localization behaviour (the sharp dip near zero field) when $V_g$ = -6 V, which indicates that the spin orbit coupling strength is different for holes and electrons. At -2.6 V, MWT1 exhibits huge magnetoresistance (> 1000% at 9 T field). When $V_g$ = 0, the device supports edge transport and the transport behaviour resembles that shown when $V_g$ = - 2.6 V. Fig. 2f shows $\ln(G_B/G_0)$ versus B/T curves measured at various temperatures with $V_g$ = 0 V. The $G_0$ is the conductance at zero field while $G_B$ is the conductance under a certain B-field. All the curves show a linear behaviour, revealing that the conductance obeys $G_B = G_0 \exp(-nB/T)$, which can be explained by the existence of a field induced Zeeman type gap (13, 14). Here, n is constant and B is the applied magnetic field. All the results in Fig. 2 agree well with previous reports and demonstrate that edge transport in monolayer $WTe_2$ has been observed. It should be emphasized that all our devices showed the same magneto-transport behaviour.

Based on the predicted results shown in Fig 1, we carried out a series of angular magnetoresistance experiments on a monolayer $WTe_2$ device (MWT6). Fig. 3a and 3b show angular dependent $R_{xx}$ at T = 10 K, $V_g$ = 4.5V when the device is rotated in-plane and out-of-plane, respectively. Based on the $V_g$ dependent resistance curve (shown in supplementary Fig S6b), this device is known to be in the edge transport state at $V_g$ = -4.5 V. The φ and θ in Fig. 3 are defined as in Fig. 1a. In the in-plane AMR measurements (Fig 3a), the applied magnetic field is always parallel to the device surface as shown in the inset of Fig. 3a, namely θ = 90°, and φ changes from 0 to 360°. In the out-of-plane measurements (Fig. 3b), φ is fixed at 0° and θ is varied from 0 to 360°. The edge channel of the monolayer $WTe_2$ is parallel to the applied magnetic field when the field is in-plane (θ = 90°). When θ = 0°, the applied magnetic field is perpendicular to the device surface. The amplitudes of AMR curves for both in-plane and out-of-plane measurements increase with increasing magnetic field and these observations agree with the theoretical predictions in Fig. 1. Fig. 3c illustrates the $V_g$ dependent amplitude of AMR for in-plane and out-of-plane AMR at T = 10 K, B = 5 T. Here, the amplitudes of the AMR are calculated as

$$\text{AMR amplitude} = \frac{\rho_{max}-\rho_{min}}{\rho_{min}} \tag{5}$$

It is clear that much higher in-plane and out of plane AMR were observed when the Fermi level was located in the band gap. When the Fermi level shifts out of the band gap, the amplitudes decrease sharply. The position of Fermi level was determined based on the $V_g$ dependent $R_{xx}$ curves. The largest amplitude of the in-plane AMR in the edge transport region is about 23%, which is an order of magnitude larger than the previously reported AMR in (~1%) 3D TIs (23), This result was anticipated based on our model of $WTe_2$. One of the fundamental ideas in the model is that the in-plane AMR in 2D topological surface states (in 3D topological insulators) and 1D edge transport channel (in QSHI) is induced by the anisotropy of the Dirac Fermi surface. The in-plane AMR of edge states in QSHI should be much larger than that of the 2D topological surface states in 3D TIs, because the tilted Fermi surface of a 2D topological surface is an ellipsoid, while it is two points for the 1D edge electrons transport in QSHI, with significantly larger anisotropy. The amplitude of the in-plane AMR in monolayer $WTe_2$ exceeds that in other ferromagnets and topological materials. The ratio of the amplitude of the in-plane AMR to the out-of-plane AMR varies from 14 – 46%. Considering that the error in the angles θ and φ in our system is less than 2° (~3%), we rule out the possibility that the in-plane AMR signal is an artefact originating from the out-of-plane AMR. Fig. 3d shows the temperature dependent AMR amplitudes for the in-plane and out-of-plane measurements. The amplitudes reach a peak at around 10 K and gradually decrease to zero with increasing temperature. In-plane AMR measurements for 3 more samples are shown in supplementary. All of them exhibit large AMR amplitudes.

Another important conclusion of the theory is that the AMR shows the smallest value when the magnetic field is parallel to the spin orientation axis and the spin orientation of the edge transport electrons should be in the yz plane (Fig. 1a), which is perpendicular to the current direction and the sample surface. To further demonstrate the validity of our theory, we fabricated a tape-shaped monolayer $WTe_2$ device (MWT7) with 2 nearly parallel edges and performed the two-terminal transport measurements. The device is shown in Fig. 4a, where the xy plane is the sample surface plane and the x axis is parallel to the device edge direction (the same as in Fig. 1a). In-plane AMR measurement was firstly carried out at 2 K under a 5 T field, as shown in Fig.4b. It is obvious that the AMR is lowest when the applied current is vertical to the applied magnetic field ($\varphi = 90^o$ and $270^o$). This demonstrates that the spin direction is in the yz-plane, because the projection of the spin on the xy plane (the sample surface) is perpendicular to the current direction. When the magnetic field is parallel to the spin projection direction, the smallest AMR value should be obtained. Once more, this result agrees with the theoretical prediction. The device was then mounted on the out-of-plane sample puck with a position where the direction of the applied current is vertical ($\varphi = 90^\circ$) to the magnetic field. By rotating the device to vary $\theta$ from $0^o$ to $360^o$ with fixed $\varphi = 90^\circ$, the out of plane AMR was measured (the black curve in Fig. 4c). From the valley points of the out-of-plane AMR curve, the spin direction in the edge channel can be confirmed. In this device, the spin points to a direction at $\theta = 149^o \pm 2^o$ and $329^o \pm 2^o$ in the yz plane, perpendicular to the applied current.

Several out-of-plane AMR curves with different φ angles in xy plane are shown in Fig.4c and these follow the same trend as the theoretical calculations in Fig.1d. This further supports our theory. Importantly, the theoretical calculations and the experiments demonstrate that AMR is an effective tool to determine the spin direction in monolayer $WTe_2$. This method is potentially suitable for other QSHIs.

In conclusion, we have developed a model to describe the relationship between the spin orientation of the helical electrons and AMR in monolayer $WTe_2$ in the QSHI states. Based on this model, extensive angular dependent magnetoresistance measurements were carried out on monolayer $WTe_2$. The material exhibited in-plane AMR amplitudes up to 22% depending on gate voltage, field and temperature. The existence of spin transport in the edge channel of monolayer $WTe_2$ has been demonstrated and the spin orientation axis of the edge channel has been determined. The experimental method applied to confirm that monolayer $WTe_2$ is a QSHI at low temperatures may now find application in probing the spin directions in the edge channels of other QSHIs. The large and tunable in-plane AMR highlights the potential use of monolayer $WTe_2$ in fabricating high sensitivity AMR logic devices.

## Methods

### Device fabrication

The $WTe_2$ single crystals were bought from HQ graphene. Monolayer $WTe_2$ nanoflakes were mechanically exfoliated onto $SiO_2$/Si wafers in a glove box with oxygen and water levels below 0.1 parts per million. The nanoflakes were examined under an optical microscope in the glove box. Clean and smooth h-BN nanoflakes were identified by atomic force microscopy and subsequently baked in the glove box before transfer. Standard e-beam lithography and sputtering were used to fabricate bottom contacts consisting of 10-15 nm Pt. After the transfer process, the device was sealed with polycarbonate and spin-coated with PMMA to prevent oxidation. Finally, PMMA on the edge of the outside electrode was removed prior to the wire-bonding process.

**Electrical and magnetic measurement**

The transport and magnetic measurements were performed in a PPMS EverCool II cryostat system (Quantum Design, San Diego, CA, USA) with a base temperature of 1.8 K and a magnetic field of up to 9 T. The PPMS horizontal rotator allows rotation along one axis. The in-plane and out-of-plane angular magnetoresistance measurements were performed on two kinds of sample pucks provided by Quantum Design. The monolayer $WTe_2$ device was sealed under a PMMA film, so transfer between the two sample pucks caused no detrimental effects.

## Acknowledgements

**General**: This research was performed in part at the RMIT Micro Nano Research Facility (MNRF) in the Victorian Node of the Australian National Fabrication Facility (ANFF) and the RMIT Microscopy and Microanalysis Facility (RMMF).

**Funding:** This research was supported by the Australian Research Council Centre of Excellence in Future Low-Energy Electronics Technologies CE170100039 (Hamilton and L. Wang), National Natural Science Foundation of China [Grants No. 11874016 (R. Q. Wang) and No. 11904107(Deng)], and by Science and Technology Program of Guangzhou (No. 2019050001).

**Author contributions:** R.-Q.W. and L.W. conceived and designed the research. C.T., S.A. and M.A. prepared the clean h-BN nanoflakes. C.T. and G.Z. prepared the bottom contacts. C.T. assembled monolayer $WTe_2$ devices, C.T. and F.X. performed the electron transport measurements. C.T., M.-X. D., A.R.H., R.-Q.W. and L.W. did the data analysis. M.-X.D., R.-Q.W. did the band theoretical modelling. C.T., M.-X. D., R.-Q.W., J.P. and L.W wrote the paper with the help from all the other co-authors.

**Competing interests:** The authors declare that they have no competing interests.

**Data and materials availability:** All data needed to evaluate the conclusions in the paper are present in the paper and/or the Supplementary Materials. Additional data related to this paper may be requested from the authors.

## Figure Legends

**Fig.1 Theoretical prediction of AMR effect in monolayer $WTe_2$. (a)** Schematic diagram of a monolayer $WTe_2$ device. The device in the upper image is tilted in the in-plane direction and the AMR is determined by φ. When the device is tilted in the yz plane, the AMR is determined by θ. **(b)** Theoretically predicted AMR curves when the device is tilted in-plane**. (c)** Theoretically predicted out-of-plane AMR curves. **(d)** Out-of-plane AMR shift at various angles between the applied current and the projection of applied magnetic field on the xy plane.

**Fig.2 Edge transport in a monolayer $WTe_2$ device (MWT1). (a)** Optical image of a monolayer $WTe_2$ device. The scale bar represents 5 μm. The few layer graphite (FLG), top h-BN and $WTe_2$ layers are labeled in the image. **(b)** $V_g$ dependent conductance at various temperatures. The dashed line represents a value of $e^2/h$. **(c-e)** Magnetoresistance at $V_g$ = 4, -2.6, -6 V at T = 2 K. **(f)** The $\ln(G_B/G_0)$ vs B/T curves derived from magnetoresistance at various temperatures at $V_g$ = 0 V. The linear curves indicate the existence of a Zeeman type gap in the edge state.

**Fig. 3 AMR measurements for a monolayer $WTe_2$ device (MWT6). (a)** In-plane AMR of MWT6 at B = 1, 5, 9 T when T = 10 K and $V_g$ = -4.5 V. The inset shows a schematic of the in-plane rotation with the orange rectangle representing the original sample surface parallel to the magnetic field (black arrow). After in-plane rotation, the device moves to the position of the yellow rectangle. **(b)** Out-of-plane AMR at B=1, 5, 9 T when T = 10 K, $V_g$= -4.5 V. The applied field is vertical to the device surface when θ = 0°, as shown in the inset. **(c)** $V_g$ dependent amplitudes of in- and out-of-plane AMR at T = 10 K, B = 5 T. **(d)** Temperature dependent

amplitudes of in- and out-of-plane AMR at $V_g$= -4.5 V, B = 5 T.

**Fig. 4 AMR measurements for a band-shaped monolayer $WTe_2$ device (MWT7). (a)** Optical image of the band-shaped device with scale bar representing 10 μm. The edge of the monolayer $WTe_2$ nanoflake is highlighted with a yellow line. The blue vertical arrow represents the applied magnetic field which is always parallel to the device surface. The x axis is parallel to the edge channel, which is at a 75° angle to the magnetic field (φ = 75°). **(b)** In-plane AMR measured at T = 2, 7, 10 K when B = 5 T, $V_g$ = -2 V. The red dashed lines represents φ = 90° and 270°. **(c)** θ dependent out-of-plane AMR with various φ values at T = 2 K, B = 5 T, $V_g$ = -2 V.

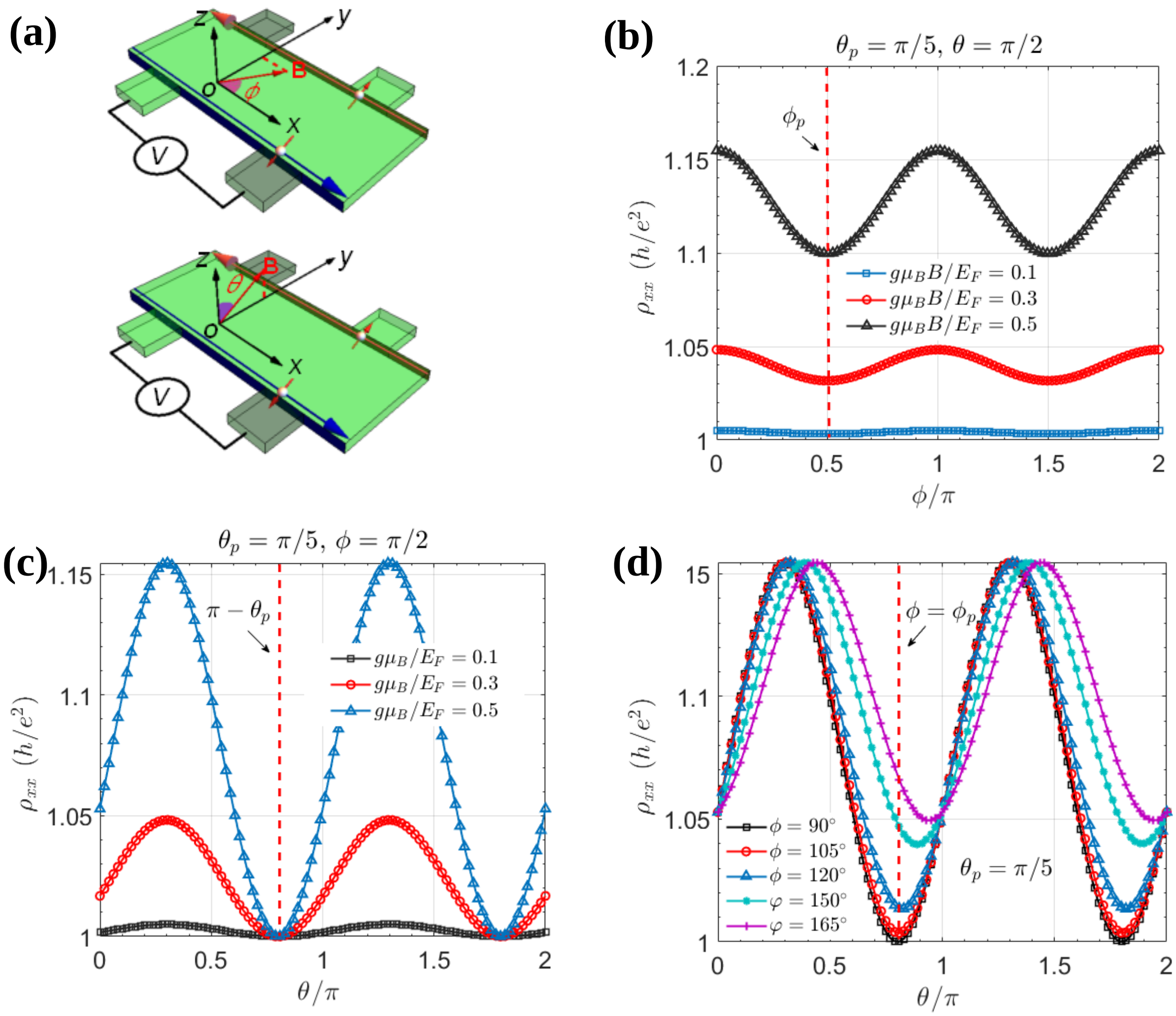

**Figure 1**

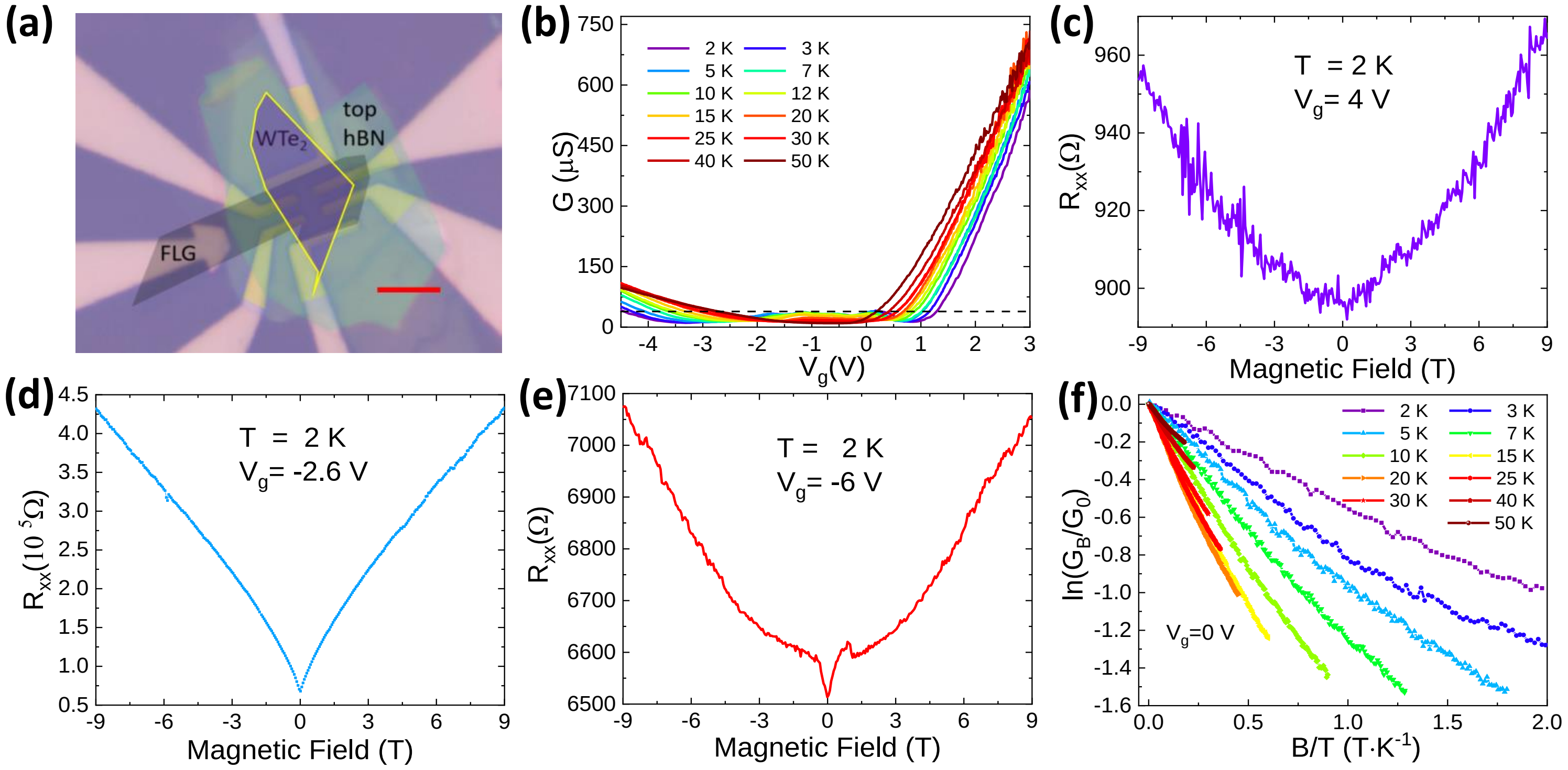

# Figure 2

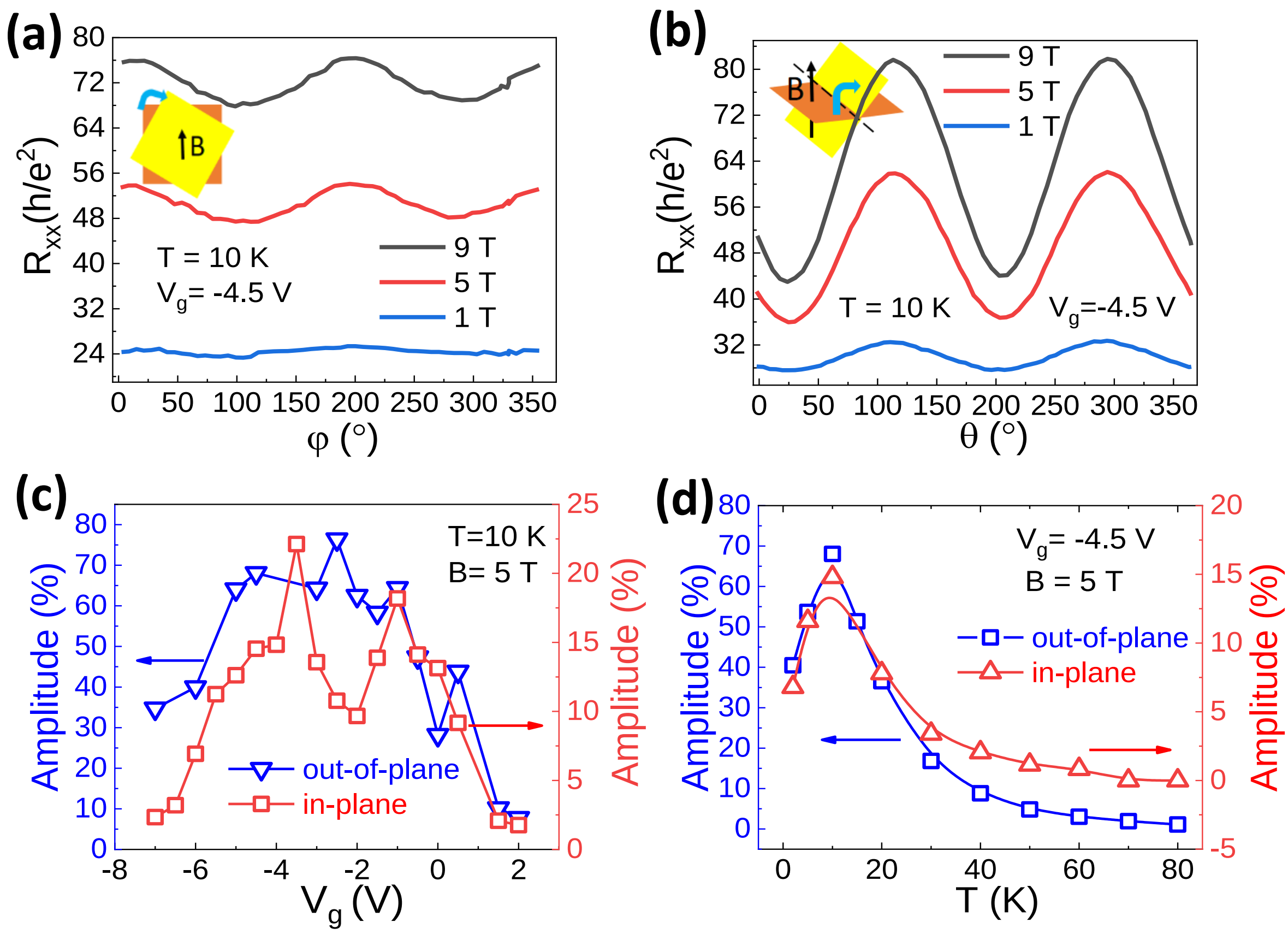

**Figure 3**

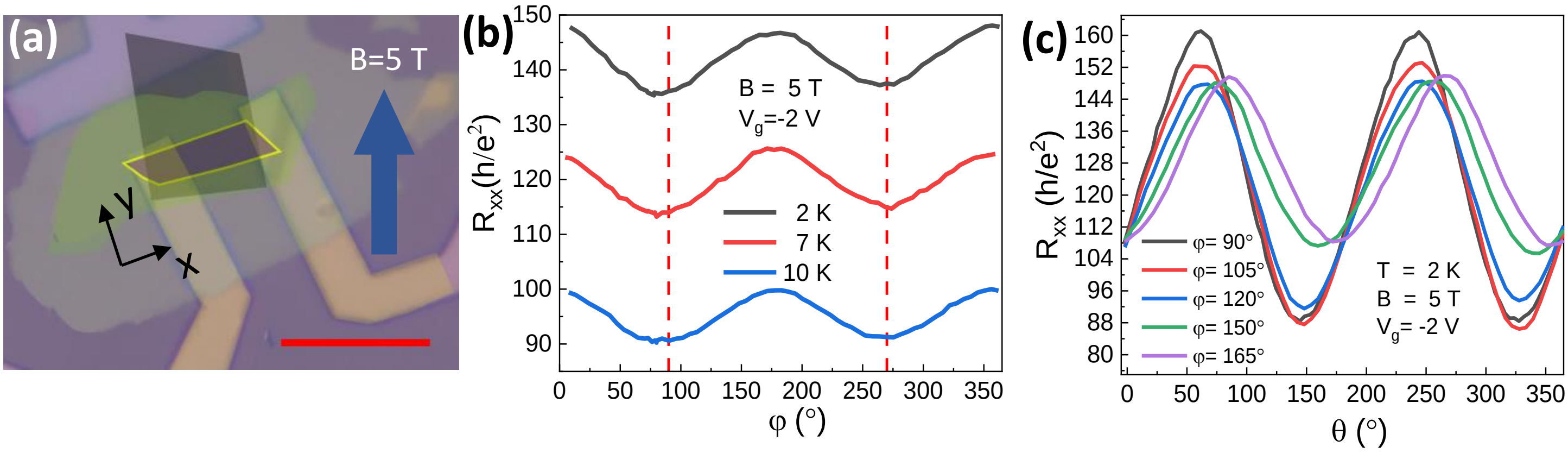

# Figure 4

# Supplementary Information for:

# Determination of the spin orientation of helical electrons in monolayer $WTe_2$

Cheng Tan[1,#], Ming-Xun Deng[2,#], Feixiang Xiang[3], Guolin Zheng[1], Sultan Albarakati[1], Meri Algarni[1], James Partridge[1], Alex R. Hamilton[3,*], Rui-Qiang Wang[2,*] and Lan Wang[1,*]

[1]School of Science and ARC Centre of Excellence in Future Low-Energy Electronics Technologies, RMIT Node, RMIT University, Melbourne, VIC 3000, Australia.
[2]Guangdong Provincial Key Laboratory of Quantum Engineering and Quantum Materials, SPTE, South China Normal University, Guangzhou 510006, China.
[3]School of Physics and ARC Centre of Excellence in Future Low-Energy Electronics Technologies, UNSW Node, University of New South Wales, Sydney, NSW 2052, Australia.

[#] These authors equally contribute to the paper.

[*] Corresponding authors. Correspondence and requests for materials should be addressed to A. H. (email: alex.hamilton@unsw.edu.au, R-Q. W. (email: wangruiqiang@m.scnu.edu.cn) and L. W. (email: lan.wang@rmit.edu.au).

**Supplementary material A: Effective Hamiltonian for the helical edge states**

As mentioned in the main text, the spin-orbit coupling induced by the vertical electric field are much weaker than the atomic spin-spin coupling, i.e.,

$$\lambda, \alpha_{x,y}, \delta_{x,z} \ll \upsilon_{x,y}, \tag{A1}$$

and so $H_1(\mathbf{k})$ can be treated as a perturbation. As the sample, placed in the $x$-$y$ plane, has two boundaries at $y = \pm\frac{L}{2}$ along the $x$ direction, the translational symmetry is broken for the $y$ direction, so that $k_y$ in the Hamiltonian needs to be repaced by an operator $k_y \to -i\partial_y$. Without perturbation, $H_0(\mathbf{k})$ is block diagonal in the spin subspace, i.e.,

$$H_0(\mathbf{k}) = \begin{pmatrix} h_\uparrow(\mathbf{k}) & 0 \\ 0 & h_\downarrow(\mathbf{k}) \end{pmatrix}, \tag{A2}$$

where

$$h_s(\mathbf{k}) = \varepsilon_{\mathbf{k}} + \begin{pmatrix} m_{\mathbf{k}} & s\hbar\upsilon_x k_x + i\hbar\upsilon_y k_y \\ s\hbar\upsilon_x k_x - i\hbar\upsilon_y k_y & -m_{\mathbf{k}} \end{pmatrix} \tag{A3}$$

with $s = +1$ $(-1)$ for the spin-$\uparrow$ (-$\downarrow$) block. Therefore, we can solve eigenvalue problem for $h_s(\mathbf{k})$ separately.

Since $k_x$ is still a good quantum number, the edge wavefunction for $h_s(\mathbf{k})$ can take the following form

$$\Phi_\kappa^s = e^{ik_x x} e^{\kappa y} \psi_s \tag{A4}$$

with $\kappa$ being real numbers. Substituting this wavefunction into the Schrodinger equation

$$h_s(k_x, -i\partial_y)\Phi_\kappa^s = E^s \Phi_\kappa^s, \tag{A5}$$

it is easy to obtain

$$\left[-c_2\kappa^2 + c_1 k_x^2 + \begin{pmatrix} m_2\left(\overline{k}^2 + \kappa^2\right) & s v_x k_x + v_y \kappa \\ s v_x k_x - v_y \kappa & -m_2\left(\overline{k}^2 + \kappa^2\right) \end{pmatrix}\right]\psi_s = \widetilde{E}_s \psi_s, \tag{A6}$$

where we have noted $v_{x,y} = \hbar\upsilon_{x,y}$, $\widetilde{E}_s = E^s - c_0$ and $\overline{k}^2 = \frac{m_0}{m_2} - \frac{m_1}{m_2}k_x^2$ for brevity. To ensure that $\psi_s$ has nontrivial solutions, i.e., $\psi_s \neq 0$, the determinant of the coefficient matrix for $\psi_s$ must equal zero, namely,

$$\left|-c_2\kappa^2 + c_1 k_x^2 - \widetilde{E}_s + \begin{pmatrix} m_2\left(\overline{k}^2 + \kappa^2\right) & s v_x k_x + v_y \kappa \\ s v_x k_x - v_y \kappa & -m_2\left(\overline{k}^2 + \kappa^2\right) \end{pmatrix}\right| = 0. \tag{A7}$$

Then, we can derive

$$\left(\widetilde{E}_s - c_1 k_x^2 + c_2\kappa^2\right)^2 - m_2^2\left(\overline{k}^2 + \kappa^2\right)^2 - v_x^2 k_x^2 + v_y^2 \kappa^2 = 0, \tag{A8}$$

from which we can solve for

$$\kappa^2 = \frac{\frac{v_y^2}{2} + \left(\widetilde{E}_s - c_1 k_x^2\right)c_2 - m_2^2\overline{k}^2}{m_2^2 - c_2^2} \pm \sqrt{\frac{v_y^2\left[\frac{v_y^2}{4} + \left(\widetilde{E}_s - c_1 k_x^2\right)c_2 - m_2^2\overline{k}^2\right] + m_2^2\left[\left(\widetilde{E}_s - c_1 k_x^2\right) - c_2\overline{k}^2\right]^2 + v_x^2\left(c_2^2 - m_2^2\right)k_x^2}{\left(m_2^2 - c_2^2\right)^2}}. \tag{A9}$$

For a given energy $\widetilde{E}_s = E$, there are four possible solutions for $\kappa$, i.e., $\kappa = \beta\kappa_\alpha$ with $\beta = \pm$, $\alpha = 1, 2$ and

$$\kappa_\alpha = \sqrt{\frac{\frac{v_y^2}{2} + \left(\widetilde{E}_s - c_1 k_x^2\right)c_2 - m_2^2\overline{k}^2}{m_2^2 - c_2^2} + (-1)^\alpha \left\{\frac{v_y^2\left[\frac{v_y^2}{4} + \left(\widetilde{E}_s - c_1 k_x^2\right)c_2 - m_2^2\overline{k}^2\right] + m_2^2\left[\left(\widetilde{E}_s - c_1 k_x^2\right) - c_2\overline{k}^2\right]^2 + v_x^2\left(c_2^2 - m_2^2\right)k_x^2}{\left(m_2^2 - c_2^2\right)^2}\right\}^{1/2}}. \tag{A10}$$

Therefore, within the sample, the general wavefunction can be expressed as a linear superposition of the four solutions

$$\Psi^s_{k_x}(r) = e^{ik_x x}\sum_{\alpha\beta} C^s_{\alpha\beta}\psi^s_{\alpha\beta}e^{\kappa_\alpha(\beta y-\frac{L}{2})}, \tag{A11}$$

where

$$\psi^s_{\alpha\beta} = \begin{pmatrix} sv_x k_x + \beta v_y\kappa_\alpha \\ F_q + (c_2 - m_2)\kappa_\alpha^2 \end{pmatrix} \tag{A12}$$

and $F_q = \widetilde{E}_s - c_1 k_x^2 - m_2\overline{k}^2$. Subsequently, the edge wavefunction can be expressed as

$$\Psi^s_{k_x}(r) = \begin{bmatrix} \begin{pmatrix} sv_x k_x + v_y\kappa_1 \\ F_q + (c_2 - m_2)\kappa_1^2 \end{pmatrix} C^s_{1+}e^{\kappa_1(y-\frac{L}{2})} + \begin{pmatrix} sv_x k_x + v_y\kappa_2 \\ F_q + (c_2 - m_2)\kappa_2^2 \end{pmatrix} C^s_{2+}e^{\kappa_2(y-\frac{L}{2})} \\ +\begin{pmatrix} sv_x k_x - v_y\kappa_1 \\ F_q + (c_2 - m_2)\kappa_1^2 \end{pmatrix} C^s_{1-}e^{-\kappa_1(y+\frac{L}{2})} + \begin{pmatrix} sv_x k_x - v_y\kappa_2 \\ F_q + (c_2 - m_2)\kappa_2^2 \end{pmatrix} C^s_{2-}e^{-\kappa_2(y+\frac{L}{2})} \end{bmatrix} e^{ik_x x}. \tag{A13}$$

The edge wavefunction should satisfy the open boundary conditions

$$\Psi^s_{k_y}(x, y = \pm\frac{L}{2}) = 0. \tag{A14}$$

Since $\mathrm{Re}(\kappa_\alpha) > 0$ and $L \to \infty$, $e^{-\kappa_\alpha L} \simeq 0$ and so the boundary condition equation can be written in a block-diagonal matrix form as

$$\begin{pmatrix} sv_x k_x + v_y\kappa_1 & sv_x k_x + v_y\kappa_2 & 0 & 0 \\ F_q + (c_2 - m_2)\kappa_1^2 & F_q + (c_2 - m_2)\kappa_2^2 & 0 & 0 \\ 0 & 0 & sv_x k_x - v_y\kappa_1 & sv_x k_x - v_y\kappa_2 \\ 0 & 0 & F_q + (c_2 - m_2)\kappa_1^2 & F_q + (c_2 - m_2)\kappa_2^2 \end{pmatrix} \begin{pmatrix} C^s_{1+} \\ C^s_{2+} \\ C^s_{1-} \\ C^s_{2-} \end{pmatrix} = 0, \tag{A15}$$

where the diagonal blocks can be combined into a $2 \times 2$ equation

$$\begin{pmatrix} sv_x k_x + \beta v_y\kappa_1 & sv_x k_x + \beta v_y\kappa_2 \\ F_q + (c_2 - m_2)\kappa_1^2 & F_q + (c_2 - m_2)\kappa_2^2 \end{pmatrix} \begin{pmatrix} C^s_{1\beta} \\ C^s_{2\beta} \end{pmatrix} = 0 \tag{A16}$$

with $\beta = +$ *for the top edge and* $\beta = -$ *for the bottom edge.* With the limitation

$$v_y\left[F_q - (c_2 - m_2)\kappa_1\kappa_2\right] = s\beta v_x k_x(c_2 - m_2)(\kappa_1 + \kappa_2), \tag{A17}$$

$C^s_{\alpha\beta}$ possess the following nontrivial solutions

$$\begin{aligned} C^s_{1\beta} &= -\left(sv_x k_x + \beta v_y\kappa_2\right), \\ C^s_{2\beta} &= +\left(sv_x k_x + \beta v_y\kappa_1\right). \end{aligned} \tag{A18}$$

At $k_x = 0$, the limitation reduces to be

$$\widetilde{E}_s - m_0 - (c_2 - m_2)\kappa_1'\kappa_2' = 0 \tag{A19}$$

with

$$\kappa_\alpha' = \sqrt{\frac{\frac{v_y^2}{2} + \widetilde{E}_s'c_2 - m_2 m_0}{m_2^2 - c_2^2} + (-1)^\alpha\left[\frac{v_y^2\left(\frac{v_y^2}{4} + \widetilde{E}_s'c_2 - m_2 m_0\right) + \left(m_2\widetilde{E}_s' - c_2 m_0\right)^2}{\left(m_2^2 - c_2^2\right)^2}\right]^{1/2}}. \tag{A20}$$

By the relation

$$\kappa_1'\kappa_2' = \sqrt{\frac{\left(\widetilde{E}_s' - m_0\right)\left(\widetilde{E}_s' + m_0\right)}{c_2^2 - m_2^2}}, \tag{A21}$$

we can derive for $\widetilde{E}'_s = m_0$ or $\widetilde{E}'_s = c_2\frac{m_0}{m_2}$. If $\widetilde{E}'_s = m_0$, $\kappa'_1\kappa'_2 = 0$ always, which can not ensure the boundary conditions $\mathrm{Re}(\kappa'_\alpha) > 0$. Therefore, $\widetilde{E}'_s = c_2\frac{m_0}{m_2}$ and $\kappa'_1\kappa'_2 = \frac{m_0}{m_2}$. Provided $m_2 > |c_2|$, we can further reduce $\kappa'_\alpha$ to be

$$\kappa'_\alpha = \frac{1}{\sqrt{m_2^2 - c_2^2}}\left[\frac{v_y}{2} + (-1)^\alpha\sqrt{\frac{v_y^2}{4} - \frac{m_0}{m_2}\left(m_2^2 - c_2^2\right)}\right]. \tag{A22}$$

Subsequently, we have

$$\psi^{s'}_{\alpha\beta}(k_x = 0) = \begin{pmatrix} \beta v_y \kappa'_\alpha \\ (c_2 - m_2)\left(\frac{m_0}{m_2} + \kappa'^2_\alpha\right) \end{pmatrix} \tag{A23}$$

and

$$\begin{aligned} C^{s'}_{1\beta} &= -\beta v_y \kappa'_2, \\ C^{s'}_{2\beta} &= +\beta v_y \kappa'_1, \end{aligned} \tag{A24}$$

such that the edge wavefunction at $k_x = 0$ is

$$\begin{aligned} \Psi^{s,\beta}_0(r) &= e^{ik_x x}\sum_\alpha C^{s'}_{\alpha\beta}\psi^{s'}_{\alpha\beta}e^{\kappa'_\alpha(\beta y - \frac{L}{2})} \\ &= \beta v_y^2 \kappa'_1\kappa'_2\sqrt{\frac{2m_2}{m_2 + c_2}}\begin{pmatrix} -\beta\sqrt{\frac{m_2+c_2}{2m_2}} \\ \sqrt{\frac{m_2-c_2}{2m_2}} \end{pmatrix}\left[e^{\kappa'_1(\beta y - \frac{L}{2})} - e^{\kappa'_2(\beta y - \frac{L}{2})}\right]e^{ik_x x}. \end{aligned} \tag{A25}$$

The edge effective Hamiltonian $H^\beta_{\mathrm{edge}}$ can be obtained by projecting $H^{\mathrm{eff}}(k_x, -i\partial_y)$ onto the basis $\{\Psi^{\uparrow,\beta}_0(r), \Psi^{\downarrow,\beta}_0(r)\}$, namely,

$$H^{\beta,ss'}_{\mathrm{edge}} = \langle\widetilde{\Psi}^{s,\beta}_0(r)|H^{\mathrm{eff}}(k_x, -i\partial_y)|\widetilde{\Psi}^{s',\beta}_0(r)\rangle \tag{A26}$$

where $\widetilde{\Psi}^{s,\beta}_0(r) = \Psi^{s,\beta}_0(r)/\sqrt{\langle\Psi^{s,\beta}_0(r)|\Psi^{s,\beta}_0(r)\rangle}$ is the normalized edge wavefunction. As we concern on properties of spin-momentum interlocking of the edge electrons, the $k_x$-dependent perturbation would dominate the effect. Therefore, we can ignore the $k_x$-independent terms in $H_1(\mathbf{k})$ and obtain the matrix elements

$$\begin{aligned} H^{\beta,\uparrow\downarrow}_1 &= \langle\widetilde{\Psi}^{\uparrow,\beta}_0(r)|\begin{pmatrix} \alpha^-_{\mathbf{k}} & i\delta_x \\ i\delta_x & \alpha^-_{\mathbf{k}} \end{pmatrix}|\widetilde{\Psi}^{\downarrow,\beta}_0(r)\rangle \simeq -i\alpha_x k_x, \\ H^{\beta,\downarrow\uparrow}_1 &= \langle\widetilde{\Psi}^{\downarrow,\beta}_0(r)|\begin{pmatrix} \alpha^+_{\mathbf{k}} & -i\delta_x \\ -i\delta_x & \alpha^+_{\mathbf{k}} \end{pmatrix}|\widetilde{\Psi}^{\uparrow,\beta}_0(r)\rangle \simeq i\alpha_x k_x. \end{aligned} \tag{A27}$$

Together with the diagonal elements

$$\begin{aligned} H^{\beta,ss'}_0 &= \langle\widetilde{\Psi}^{s,\beta}_0(r)|H_0(k_x, -i\partial_y)|\widetilde{\Psi}^{s,\beta}_0(r)\rangle \\ &= \left[c_0 + \frac{c_2}{m_2}m_0 + \left(c_1 - \frac{m_1}{m_2}c_2\right)k_x^2 - s\beta\hbar\upsilon_x\sqrt{\frac{m_2^2 - c_2^2}{m_2^2}}k_x\right]\delta_{ss'}, \end{aligned} \tag{A28}$$

the edge effective Hamiltonian can be expressed in a matrix form as

$$H^\beta_{\mathrm{edge}} = E_0 + \gamma k_x^2 + \hbar k_x\begin{pmatrix} -\upsilon_F & -i\upsilon_\alpha \\ i\upsilon_\alpha & \upsilon_F \end{pmatrix}, \tag{A29}$$

where the parameters above are defined as $E_0 = c_0 + \frac{m_0}{m_2}c_2$, $\gamma = c_1 - m_1c_2/m_2$, $\upsilon_\alpha = \alpha_x/\hbar$ and $\upsilon_F = \beta\upsilon_x\sqrt{1 - c_2^2/m_2^2}$. The constant term is trivial and can be omitted, such that Eq. (A29) reduces to Eq. (3) of the main text.

**Supplementary material B: Resistivity for the edge channel**

The charge current density is given by

$$j_x = \frac{e}{L_x L_y} \sum_{k_x, \eta=\pm} \langle \psi^\eta_{k_x} | \widehat{\upsilon}_x | \psi^\eta_{k_x} \rangle \delta f(\varepsilon^\eta_{k_x}), \tag{A30}$$

where $\psi^\eta_{k_x}$ are eigenstates for $H^\beta_{\text{edge}}$, $\widehat{\upsilon}_x = \partial H^\beta_{\text{edge}} / (\hbar \partial k_x)$ and $\delta f(\varepsilon^\eta_{k_x})$ stands for external-field-induced shift of the distribution function. In the relaxation time approximation, $\delta f(\varepsilon^\eta_{k_x})$ can be calculated by the Boltzmann kinetic equation, which reads as

$$\delta f(\varepsilon^\eta_{k_x}) = eE \langle \psi^\eta_{k_x} | \widehat{\upsilon}_x | \psi^\eta_{k_x} \rangle \frac{\partial f_0(\varepsilon^\eta_{k_x})}{\partial \varepsilon^\eta_{k_x}} \tau_a. \tag{A31}$$

Here, $f_0(x) = \left[1 + \exp[\left(\frac{x - E_F}{k_B T}\right)]\right]^{-1}$ with $E_F$ being the Fermi energy is the Fermi-Dirac distribution function, $\tau_a$ denotes the relaxation time and $E$ represents magnitude of the applied electric field. Substituting $\delta f(\varepsilon^\eta_{k_x})$ into Eq. (A30) yields

$$j_x = -\frac{e^2}{h} \frac{\tau_a}{L_y} E \sum_{\eta=\pm} \int d\varepsilon^\eta_{k_x} | \langle \psi^\eta_{k_x} | \widehat{\upsilon}_x | \psi^\eta_{k_x} \rangle | \frac{\partial f_0(\varepsilon^\eta_{k_x})}{\partial \varepsilon^\eta_{k_x}}. \tag{A32}$$

When a magnetic field $\mathbf{B} = (B_x, B_y, B_z) = B(\sin\theta\cos\phi, \sin\theta\sin\phi, \cos\theta)$ is applied, the Hamiltonian becomes

$$H^\beta_{\text{edge}} = \gamma k_x^2 + \hbar k_x \vec{\sigma} \cdot \boldsymbol{n} - g\mu_B (\vec{\sigma} \cdot \mathbf{B}). \tag{A33}$$

For convenience of calculation, we rotate the spin quantization axis to align with $\boldsymbol{n} = \left(0, \sin\theta_p, -\cos\theta_p\right)$ by performing a unitary transform on the Hamiltonian and obtain

$$\widetilde{H}^\beta_{\text{edge}} = U^{-1} H^\beta_{\text{edge}} U = \hbar \widetilde{\upsilon}_F k_x \sigma_z - \widetilde{\mathbf{M}} \cdot \vec{\sigma} \tag{A34}$$

with $U = \sin\frac{\theta_p}{2} + i\cos\frac{\theta_p}{2}\sigma_x$, where $\widetilde{\upsilon}_F = \sqrt{\upsilon_\alpha^2 + \upsilon_F^2}$, $\widetilde{M}_x = M_x$,

$$\begin{aligned} \widetilde{M}_y &= M_y \cos\theta_p + M_z \sin\theta_p, \\ \widetilde{M}_z &= M_y \sin\theta_p - M_z \cos\theta_p, \end{aligned} \tag{A35}$$

and $\mathbf{M} = g\mu_B \mathbf{B}$. Then, the dispersion can be easily obtained as

$$\varepsilon^\eta_{k_x} = \gamma k_x^2 + \eta \sqrt{\left(\hbar \widetilde{\upsilon}_F k_x - \widetilde{M}_z\right)^2 + M_x^2 + \widetilde{M}_y^2} \tag{A36}$$

with $\eta = \pm$ for the conduction and velance band. The corresponding eigenstate is

$$\psi^\eta_{k_x} = \frac{1}{\sqrt{2}} \begin{pmatrix} -\sqrt{1 + \frac{\hbar\widetilde{\upsilon}_F k_x - \widetilde{M}_z}{\eta\sqrt{\left(\hbar\widetilde{\upsilon}_F k_x - \widetilde{M}_z\right)^2 + \left(M_x^2 + \widetilde{M}_y^2\right)}}} e^{i\varphi_M} \\ \eta\sqrt{1 - \frac{\hbar\widetilde{\upsilon}_F k_x - \widetilde{M}_z}{\eta\sqrt{\left(\hbar\widetilde{\upsilon}_F k_x - \widetilde{M}_z\right)^2 + \left(M_x^2 + \widetilde{M}_y^2\right)}}} \end{pmatrix} \tag{A37}$$

with $\varphi_M = \tan^{-1}\left(\widetilde{M}_y / M_x\right)$. Therefore, we can further derive

$$\langle \psi^\eta_{k_x} | \widehat{\upsilon}_x | \psi^\eta_{k_x} \rangle = 2\frac{\gamma}{\hbar} \frac{\widetilde{M}_z}{\hbar\widetilde{\upsilon}_F} + \widetilde{\upsilon}_F \frac{\hbar\widetilde{\upsilon}_F k_x}{\sqrt{(\hbar\widetilde{\upsilon}_F k_x)^2 + M_x^2 + \widetilde{M}_y^2}} \tag{A38}$$

and then

$$j_x = -\frac{e^2}{h} \frac{\widetilde{\upsilon}_F \tau_a}{L_y} E \sum_{\eta=\pm} \int d\varepsilon^\eta_{k_x} \left| \frac{\hbar\widetilde{\upsilon}_F k_x}{\sqrt{(\hbar\widetilde{\upsilon}_F k_x)^2 + M_x^2 + \widetilde{M}_y^2}} + 2\frac{\gamma}{\hbar\widetilde{\upsilon}_F} \frac{\widetilde{M}_z}{\hbar\widetilde{\upsilon}_F} \right| \frac{\partial f_0(\varepsilon^\eta_{k_x})}{\partial \varepsilon^\eta_{k_x}}. \tag{A39}$$

At low temperatures, we can approximate $\frac{\partial f_0(\varepsilon^{\eta}_{k_x})}{\partial \varepsilon^{\eta}_{k_x}} = -\delta\left(\varepsilon^{\eta}_{k_x} - E_F\right)$. By using property of the $\delta$ function, we can obtain

$$j_x = \frac{e^2}{h}\frac{\widetilde{\upsilon}_F \tau_a}{L_y} E \sum_{\chi=\pm} \left| \frac{\hbar\widetilde{\upsilon}_F k_\chi}{\sqrt{\left(\hbar\widetilde{\upsilon}_F k_\chi\right)^2 + M_x^2 + \widetilde{M}_y^2}} + 2\frac{\gamma}{\hbar\widetilde{\upsilon}_F}\frac{\widetilde{M}_z}{\hbar\widetilde{\upsilon}_F} \right|, \tag{A40}$$

where $k_\chi$, solved from the equation $E_F = \gamma k_\chi^2 + \eta\sqrt{\left(\hbar\widetilde{\upsilon}_F k_\chi - \widetilde{M}_z\right)^2 + M_x^2 + \widetilde{M}_y^2}$ is given by

$$k_\pm = \left( \pm\sqrt{\frac{1}{\left(1-\xi^2\right)^2} - \frac{M_x^2 + \widetilde{M}_y^2}{E_F^2\left(1-\xi^2\right)}} - \frac{\xi}{1-\xi^2} \right) \frac{E_F}{\hbar\widetilde{\upsilon}_F} \tag{A41}$$

with $\xi = 2\frac{\gamma}{\hbar\widetilde{\upsilon}_F}\frac{\widetilde{M}_z}{\hbar\widetilde{\upsilon}_F}$. The resistivity is defined by $\rho_{xx}(\theta, \phi) \equiv E/j_x$, which results in Eq. (5) of the main text.

**Supplementary material C: Transport data of monolayer $WTe_2$ devices**

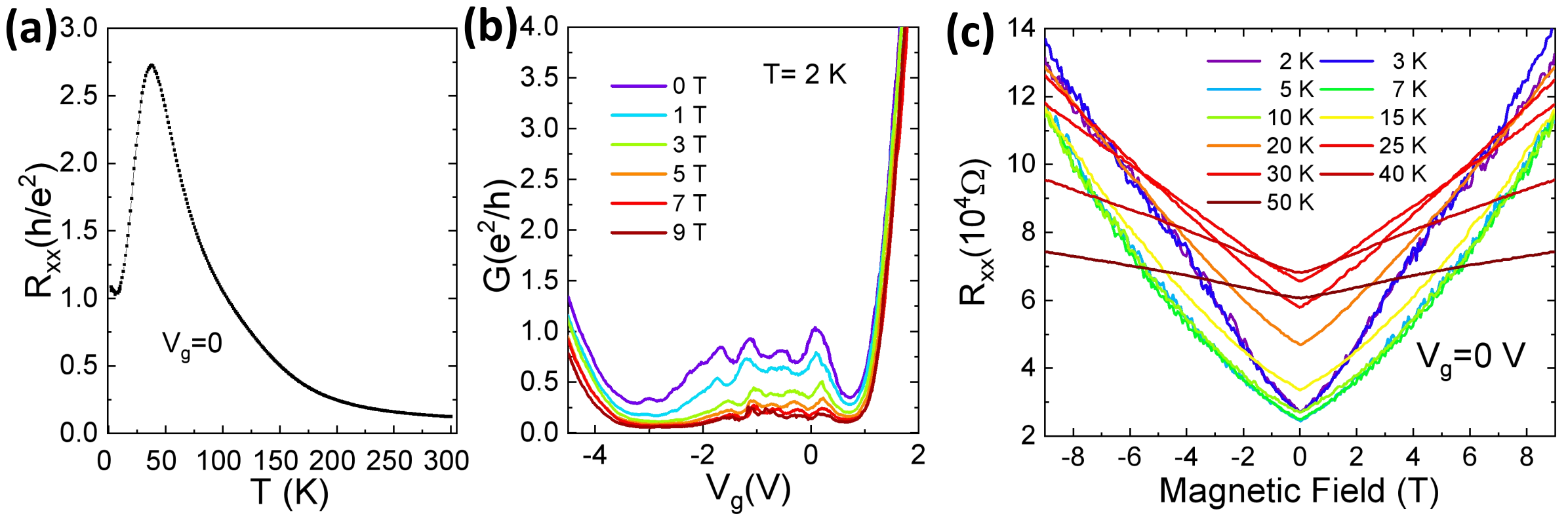

**Supplementary Fig.S1 Transport data for MWT1.** The top h-BN thickness=11.6 nm, channel length = 1.1 μm. The channel length is defined by the spacing between the 2 Pt probes (exclude source and drain probes) in longitudinal direction (same as below). **(a)** Temperature dependent $R_{xx}$ curve at B=0 T, $V_g$= 0 T. **(b)** Gate voltage dependent conductance (G) curves from 0 T to 9 T at T=2 K. **(c)** Magnetoresistance curves from 2 K to 50 K at $V_g$=0 V. In **(b)** and **(c)**, the magnetic field is perpendicular to the device surface.

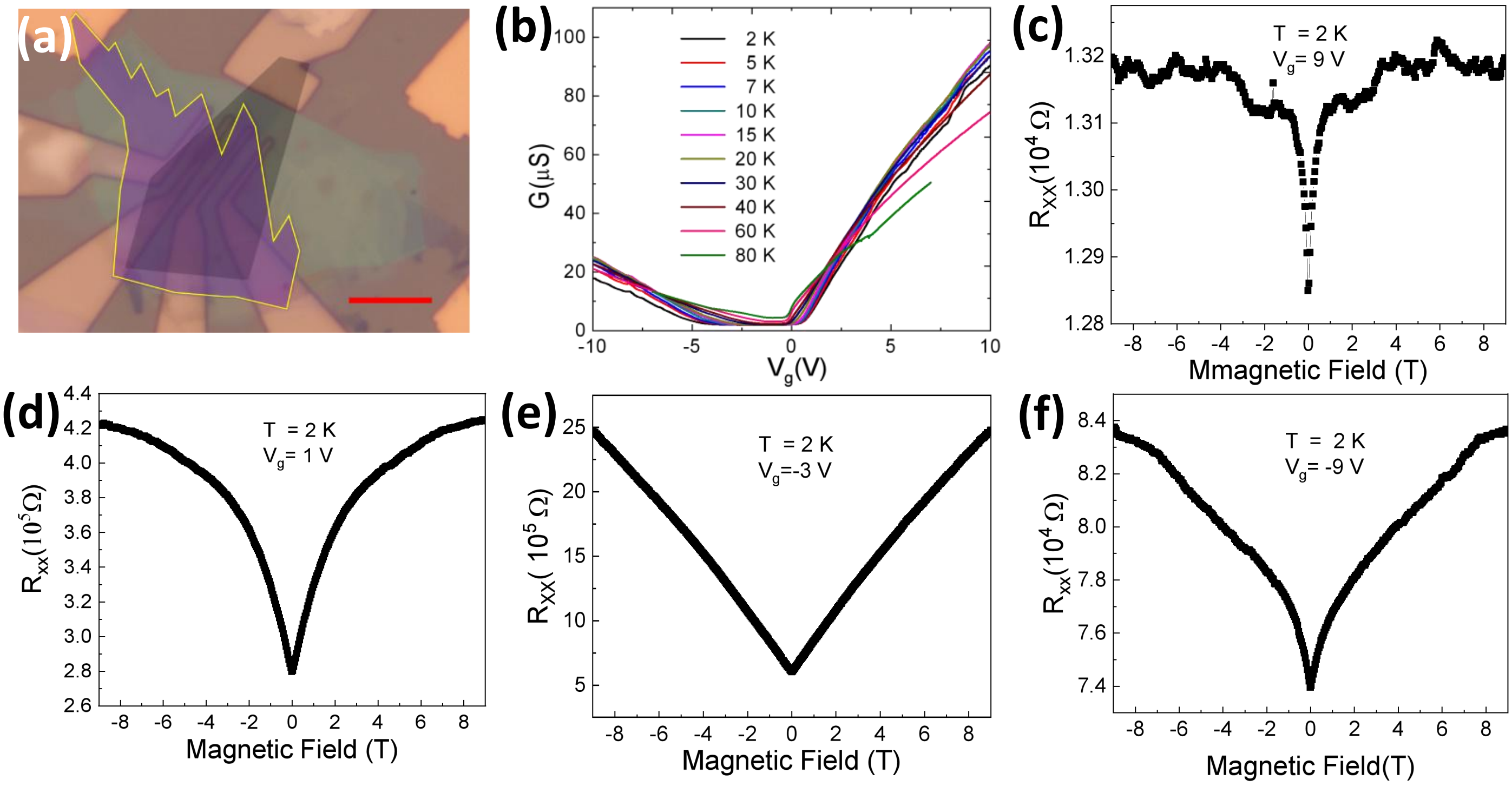

**Supplementary Fig.S2 Transport data for MWT2.** The top h-BN thickness=19 nm, channel length= 1.8 μm. **(a)** Optical image of device, the scale bar represents 10 μm. The shape of FLG (black), top h-BN (green) and monolayer $WTe_2$ (yellow surrounded purple) are highlighted in the picture. **(b)** Gate voltage dependent G curves from 2 K to 80 K at B=0 T. **(c-f)** Magnetoresistance curves at $V_g$= 9,1,-3,-9 V, respectively. All curves are measured at 2 K. The MR curves exhibit a similar evolution to those of MWT1.

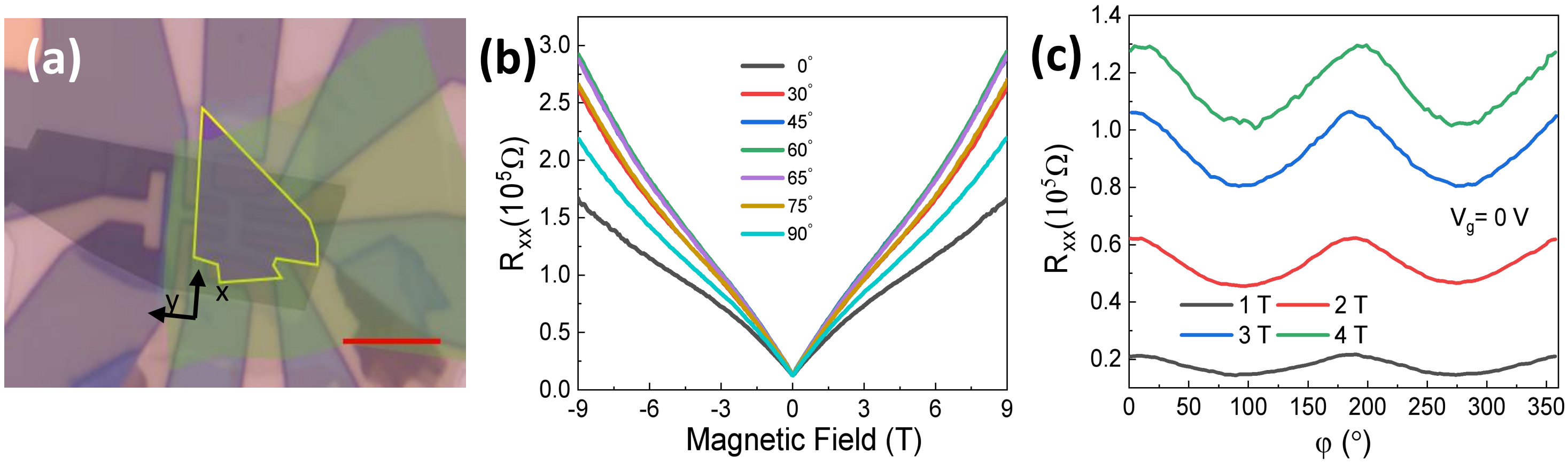

**Supplementary Fig.S3 Transport data for MWT3.** The top hBN thickness=12 nm, channel length= 1 μm. The gate for this sample didn't work, the device was tested only at $V_g$=0 V. **(a)** Optical image of device, the scale bar represents 10 μm. The shape of FLG (black), top hBN (green) and monolayer $WTe_2$ (yellow surrounded purple) are highlighted in the picture. **(b)** Field dependent $R_{xx}$ curves at different out-of-plane tilt angles. At 90° the device surface and the tested $WTe_2$ edge are parallel to the magnetic field. **(c)** φ dependent inplane AMR curves from 1T to 4 T, the xy plane is defined in **(a).** The $R_{xx}$ channel which is close to the gate electrode was measured.

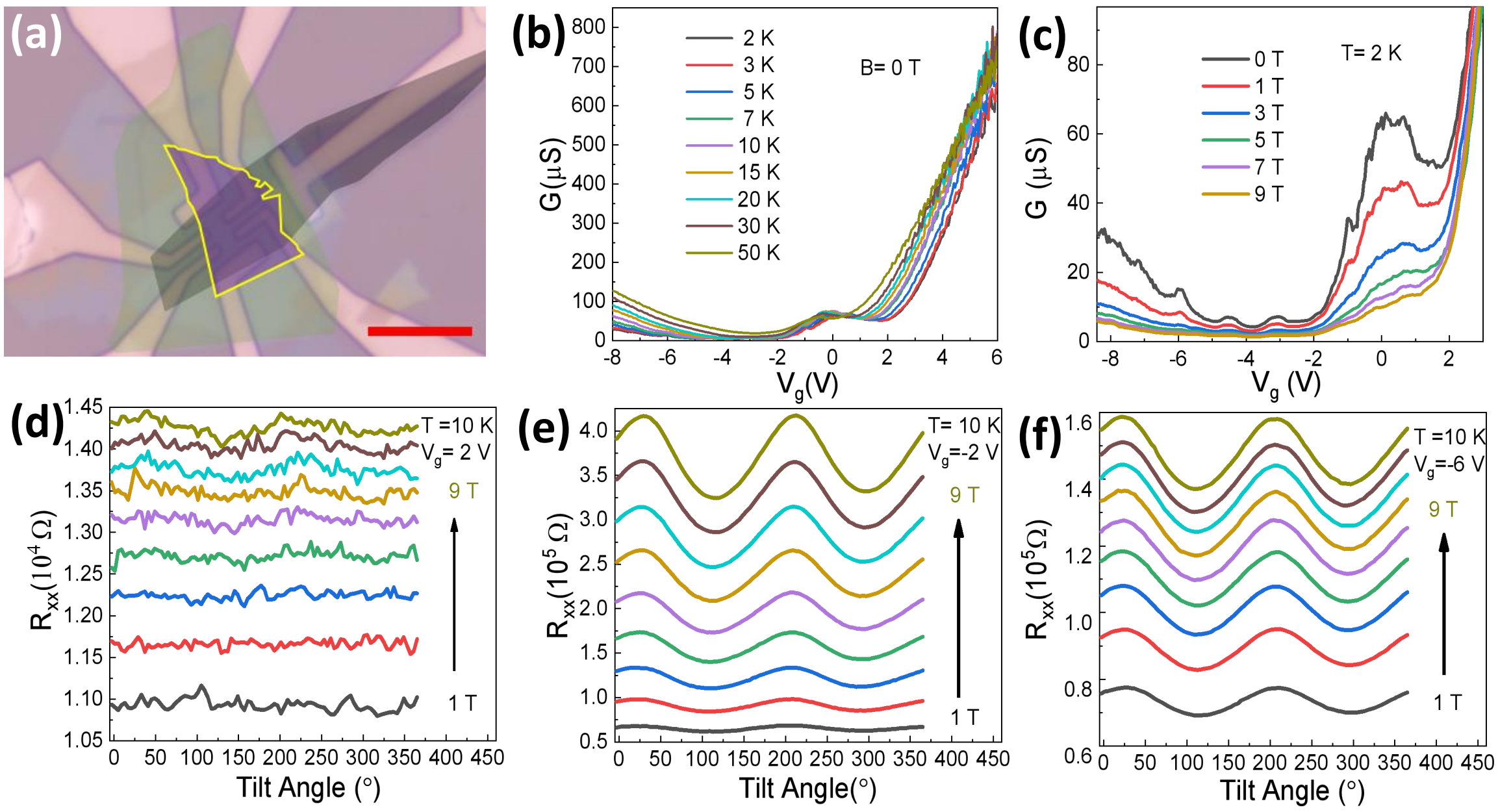

**Supplementary Fig.S4 Transport data for MWT4.** The top h-BN thickness=15 nm, channel length= 1 μm. **(a)** Optical image of device, the scale bar represents 10 μm. The shape of FLG (black), top h-BN (green) and monolayer $WTe_2$ (yellow surrounded purple) are highlighted in the picture. **(b)** Gate voltage dependent G curves from 2 K to 50 K at B=0 T. **(c)** Gate voltage dependent G curves from 0 T to 9 T at T=2 K. **(d-f)** Tilt angle dependent inplane AMR curves from 1 T to 9 T at T=10 K with $V_g$= 2, -2, -6 V, respectively. As the tested edge here (the edge channel close to the gate electrode) is not straight, the xy plane is hard to define. From the figures, it is obvious that the inplane AMR amplitudes can be tuned by the gate voltage.

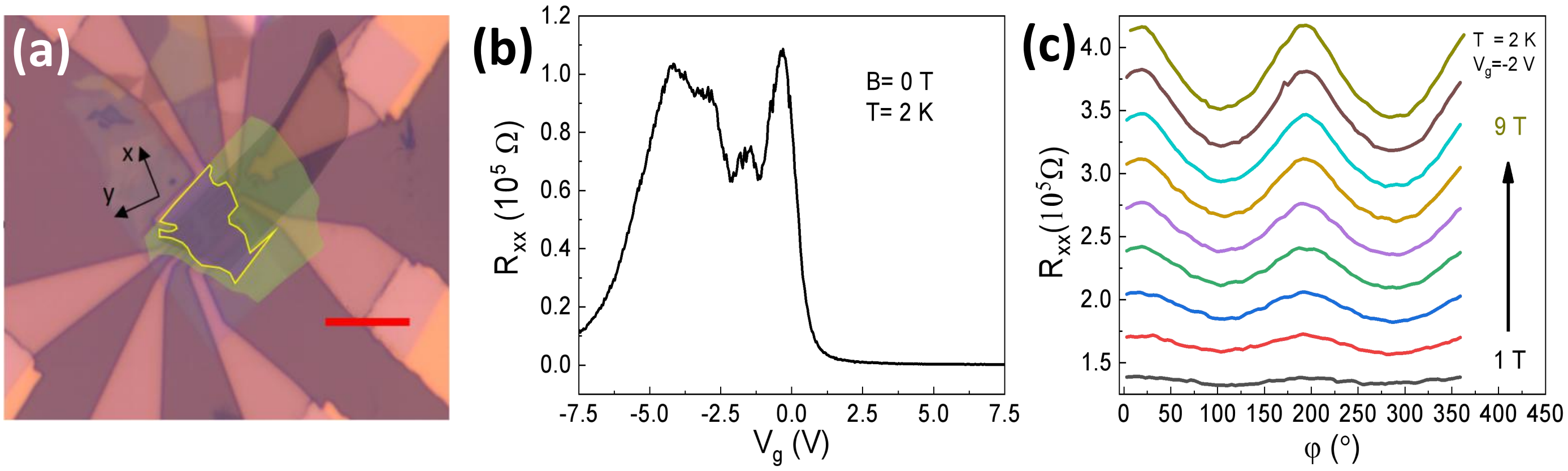

**Supplementary Fig.S5 Transport data for MWT5.** The h-BN thickness=12.3 nm, channel length= 1.2 μm. **(a)** Optical image of device, the scale bar represents 10 μm. The shape of FLG (black), top h-BN (green) and monolayer $WTe_2$ (yellow surrounded purple) are highlighted in the picture. **(b)** Gate voltage dependent $R_{xx}$ curve at T=2 K, B=0 T. **(c)** φ dependent $R_{xx}$ curves from 1 T to 9 T at T= 2 K, $V_g$= -2 V. The xy plane is defined in **(a)**. The $R_{xx}$ channel close to the gate electrode was measured.

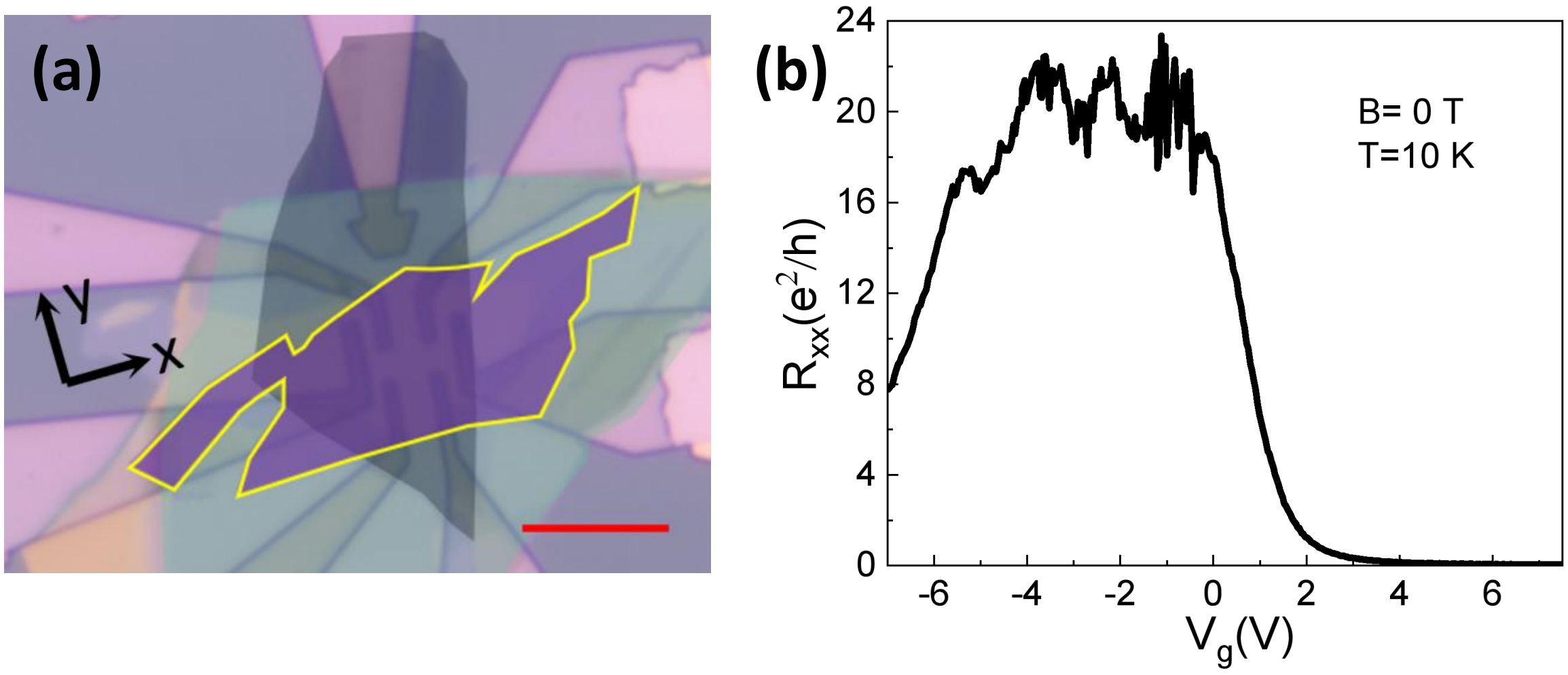

**Supplementary Fig.S6 Transport data for MWT6.** The h-BN thickness=20.6 nm, channel length= 2.5 μm.**(a)** Optical image of device, the scale bar represents 10 μm. The shape of FLG (black), top h-BN (green) and monolayer $WTe_2$ (yellow round purple) are highlighted in the picture. The xy plane is defined by the measured edge channel which is away from the gate electrode. **(b)** Gate voltage dependent $R_{xx}$ curve at T=10 K, B=0 T.

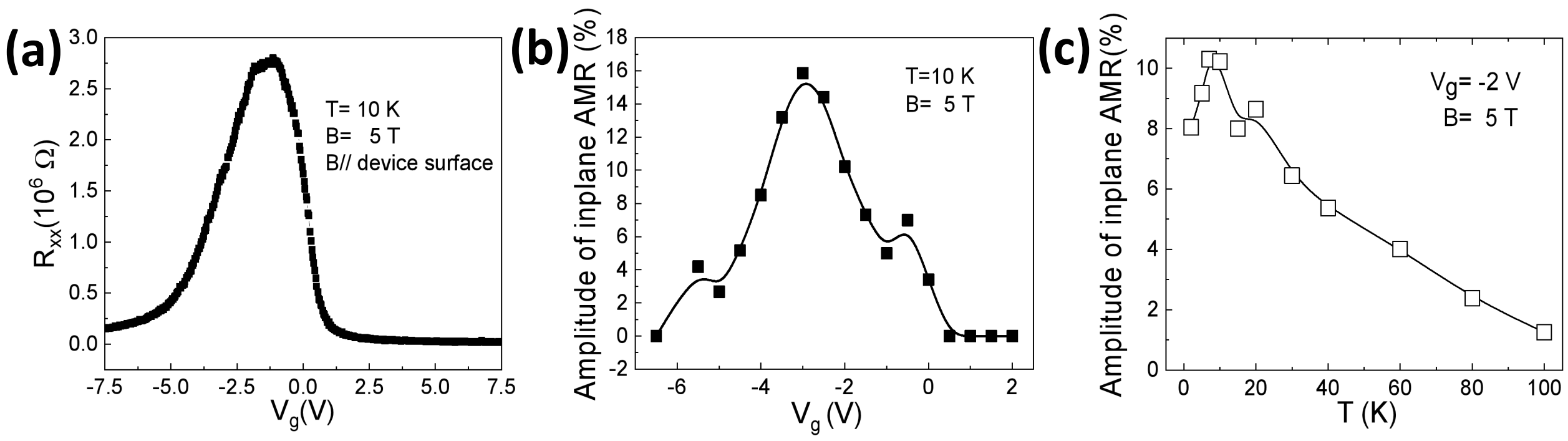

**Supplementary Fig.S7 Transport data for MWT7.** The top h-BN thickness=14.6 nm, channel length= 4.6 μm. **(a)** Gate voltage dependent $R_{xx}$ curve at T=10 K, B=5 T. This curve is measured when the measured channel is parallel to the magnetic field (inplane magnetic field). **(b)** Gate voltage dependent inplane AMR amplitudes at T=10 K, B=5 T. **(c)** Temperature dependent inplane AMR amplitudes at B=5 T, $V_g$=-2 V.